\documentclass[12pt,a4paper]{article}

\usepackage{a4}

\usepackage{amsmath}
\usepackage{amsfonts}
\usepackage{amssymb}
\usepackage{multicol}  

\usepackage{hyperref}              
\hypersetup{pdftitle={F-theory Uplifts and GUTs}, 
    pdfauthor={Ralph Blumenhagen, Thomas W. Grimm, Benjamin Jurke and Timo Weigand}, 
    pdfsubject={String theory model building},
    bookmarksopen, bookmarksnumbered, bookmarksopenlevel=2}

\newcommand{\beq}{\begin{equation}}  \newcommand{\eeq}{\end{equation}}
\newcommand{\bal}{\begin{aligned}}   \newcommand{\eal}{\end{aligned}}

\def\ov{\overline}

\def\IP{\mathbb{P}}

\begin{document}

\baselineskip=14pt
\parskip 5pt plus 1pt

\vspace*{-1.5cm}
\begin{flushright}    
  {\small
  MPP-2009-66\\
  SLAC-PUB-13654
  }
\end{flushright}

\vspace{2cm}
\begin{center}        
  {\LARGE
  F-theory Uplifts and GUTs 
  }
\end{center}

\vspace{0.75cm}
\begin{center}        
  Ralph Blumenhagen$^{1}$, Thomas W.~Grimm$^{2}$, \\[0.1cm]
  Benjamin Jurke$^{1}$ and Timo Weigand$^{3}$
\end{center}

\vspace{0.15cm}
\begin{center}        
  \emph{$^{1}$ Max-Planck-Institut f\"ur Physik, F\"ohringer Ring 6, \\ 
               80805 M\"unchen, Germany}
  \\[0.15cm]
  \emph{$^{2}$ Bethe Center for Theoretical Physics and \\ 
               Physikalisches Institut der Universit\"at Bonn, Nussallee 12, \\ 
               53115 Bonn, Germany}
  \\[0.15cm]
  \emph{$^{3}$ SLAC National Accelerator Laboratory, Stanford University, \\
               2575 Sand Hill Road, Menlo Park, CA 94025, USA}
\end{center} 

\vspace{2cm}


\begin{abstract}
We study the F-theory uplift of Type IIB orientifold models on compact Calabi-Yau threefolds containing divisors which are del Pezzo surfaces. We consider two examples defined via del Pezzo transitions of the quintic.  The first model has an orientifold projection leading to two disjoint O7-planes and the second involution acts via an exchange of two del Pezzo surfaces. The two uplifted fourfolds are generically singular with minimal gauge enhancements over a divisor and, respectively, a curve in the non-Fano base. We study possible further degenerations of the elliptic fiber leading to F-theory GUT models based on subgroups of $E_8$. 
\end{abstract}

\clearpage


\section{Introduction}

Grand unified theories (GUTs) provide a beautiful field theoretic proposal for physics beyond the weak scale. It is thus natural to investigate whether one can embed supersymmetric GUTs into string theory. This has been a challenging question throughout the history of string theory. It was noticed as early as in the mid eighties that  the ten-dimensional perturbative $E_8\times E_8$ heterotic string naturally incorporates GUT gauge groups like $SU(5)$ and $SO(10)$ upon compactification to four flat dimensions together with the desired matter particle content and (Yukawa) couplings. Finding a completely realistic string model, though, turned out to be not that straightforward, despite tremendous progress over the years, as summarized e.g. in \cite{Nilles:2008gq}. On the other hand, for the heterotic string there is no natural origin for the small hierarchy $M_X/M_{\rm pl}\simeq 10^{-3}$ and one has to invoke large threshold corrections or anisotropic backgrounds for its explanation. 

Alternatively to the heterotic string, since the mid nineties D-brane models of various kinds have been discussed as candidate stringy realizations of the MSSM. These constructions go under the name of orientifold models, see e.g. \cite{Blumenhagen:2005mu,Blumenhagen:2006ci,Marchesano:2007de} for more recent reviews. In this construction it was observed \cite{Blumenhagen:2001te} that $SO(10)$ and $SU(5)$ GUTs were obstructed by the perturbative absence of matter fields in the ${\bf 16}$ representation of $SO(10)$ and by the absence of the top Yukawa coupling ${\bf 10}\, {\bf 10}\, {\bf 5}_{{\bf H}}$ for the $SU(5)$ case. 

More recently it has been realized that the aforementioned problems with realizing simple GUT groups in orientifold constructions are nicely reconciled in F-theory models on elliptically-fibered Calabi-Yau
fourfolds \cite{Donagi:2008ca, Beasley:2008dc, Beasley:2008kw, Donagi:2008kj}.

Due to the strong backreaction, only in a global $g_s\to 0$ limit a general F-theory model is expected to correspond to a Type IIB Calabi-Yau orientifold with D-branes. More generally, F-theory inherently captures features which are non-perturbative in $g_s$. These allow for the possibility of non-perturbative gauge enhancements and the appearance of exceptional groups $E_6$, $E_7$, $E_8$ in F-theory. By a further unfolding of these exceptional groups it is also possible to realize the spinor representation of a GUT $SO(10)$ as well as the top-quark Yukawa couplings ${\bf 10}\, {\bf 10}\, {\bf 5}_{{\bf H}}$ in 
GUT $SU(5)$. For four-dimensional models, the basis $B$ is a threefold and the 7-branes wrap complex surfaces.

To suppress gravity (bulk) induced effects on the brane physics, it was proposed in \cite{Beasley:2008kw} that a decoupling limit of gravity should in principle exist. This implies that the GUT physics should be localized on a 7-brane wrapping a shrinkable four-cycle in the base of the elliptically-fibered Calabi-Yau fourfold. Such shrinkable surfaces are given by del Pezzo surfaces $dP_n$, which are $\IP^2$ blown-up at $n=0,\ldots,8$ different points and $\IP^1\times\IP^1$. It was further proposed to break the GUT symmetry to the Standard Model by means of a non-vanishing $U(1)_Y$ gauge flux \cite{Beasley:2008kw, Donagi:2008kj}. For the hypercharge to remain massless, this flux must be supported on two-cycles in the del Pezzo surface which are trivial as two-cycles in the base \cite{Buican:2006sn}. Therefore, the existence of del Pezzo surfaces with such 'trivial' two-cycles is the starting point for a concrete implementation of these ideas in compact Calabi-Yau fourfolds. Of course, the realization of three generations, realistic Yukawa textures, suppressed proton decay and a solution to the doublet-triplet splitting problem imposes more conditions on the fourfold geometry and the four-form fluxes on them.  Recent studies of these and further phenomenological questions and of the associated model building prescriptions in local constructions include \cite{Heckman:2008es, Marsano:2008jq, Heckman:2008qt, Font:2008id, Heckman:2008qa, Blumenhagen:2008aw, Heckman:2008jy, Bourjaily:2009vf, Hayashi:2009ge, Heckman:2009bi, Bouchard:2009bu, Randall:2009dw, Heckman:2009de, Tatar:2009jk, Jiang:2009za, Li:2009cy}. Progress towards compact models has been made in \cite{Andreas:2009uf,Donagi:2009ra,Marsano:2009ym}.

As an intermediate step, in \cite{Blumenhagen:2008zz} it was analyzed to what degree all these geometric conditions can be met already in Type IIB orientifolds. Clearly, such perturbative models face the problem of generating all genuine $E_8$ structures at best non-perturbatively, such as the ${\bf 10}\,{\bf 10}\, {\bf 5}_{\bf H}$ Yukawa coupling \cite{Blumenhagen:2007zk}. Nevertheless they provide a good starting point in several respects: The stringy consistency conditions such as flux quantization, tadpole cancellation, D-term supersymmetry conditions must all show up in an analogous way in F-theory. For instance, the GUT symmetry breaking via $U(1)_Y$ flux on del Pezzo surfaces also works in orientifold models \cite{Blumenhagen:2008zz,Blumenhagen:2008aw}. Second, the orientifolded Calabi-Yau threefold geometries might lead to interesting fourfold geometries once one understands their uplift. Such an approach was initiated in \cite{Collinucci:2008zs}, where it was analyzed how simple orientifolds give rise to Calabi-Yau fourfolds.

It is the aim of this paper to analyze this uplifting further and generalize it in particular to the orientifolds of \cite{Blumenhagen:2008zz}, which are guaranteed to contain del Pezzo surfaces with trivial two-cycles. This goes beyond the analysis of \cite{Collinucci:2008zs}, as the orientifolds of major interest have either O7-planes with more than one components or are defined via exchange involutions. In the first case, the del Pezzo is one component of the orientifold locus and in the second case the del Pezzo is exchanged with a mirror del Pezzo. We will discuss one example of each kind in detail and list the results of the uplift for many more examples in an appendix.

The idea is to first understand how these geometries lift to elliptically fibered Calabi-Yau fourfolds. Clearly, in the Sen limit \cite{Sen:1997gv} these fourfolds must reproduce the location of the former orientifold planes. We will see that these uplifts generically have singularities, i.e.~they are Weierstrass fibrations over non-Fano threefolds. The presence of these singularities is linked to the existence of a minimal non-Abelian gauge symmetry in the orientifold model. The latter is essentially a consequence of having branes on del Pezzo divisors which cannot be deformed. 

As we will explicitly verify, the uplifted fourfolds have complex structure moduli which do not have an orientifold analogue and therefore allow for more general degenerations of the elliptic fiber. In particular, we are interested in realizing the $E_8$ structures on these fourfolds. We will show at  two concrete prototype examples that indeed, even though the starting point was an orientifold model, the fourfold allows for $E_8$-type degeneration. However, it turns out that there are still some restrictions on the GUT structures.

\section{Preliminaries on Type IIB orientifolds}

In this section we collect some aspects of Type IIB Calabi-Yau orientifold compactifications with space-time filling D7-branes which are relevant for our discussion. More details can be found e.g. in \cite{Blumenhagen:2008zz}. We consider an orientifold projection which allows for O3 and O7-planes and takes the form $(-1)^{F_L} \Omega \sigma$.  Here $\sigma$ is a holomorphic and isometric involution of an internal Calabi-Yau threefold $X$. This involution splits the cohomology groups $H^p(X)$ and homology groups $H_q(X)$ into positive and negative eigenspaces $H^p_\pm(X)$ and $H^\pm_q(X)$. In particular, this split can be used to decompose the triple intersection form for divisors of $X$. Due to the invariance of the volume form one finds that three elements in $ D^-_i \in H_4^-(X)$ as well as two elements of $D_i^+ \in H_4^+(X)$ and one of $H_4^-(X)$ do not intersect \cite{Grimm:2004uq} 
\beq \label{pmvanish} 
  D^-_1 D^-_2 D^-_3 = 0\, \qquad D^-_1 D_2^+ D_3^+ = 0.
\eeq

In order to cancel tadpoles, the orientifold model has to include a set of D7-branes which fill (four-dimensional) space-time and wrap holomorphic four-cycles $D_a$ of the Calabi-Yau manifold. The orientifold symmetry $\sigma$ maps $D_a$ to its orientifold image $D_a'$ so that in the upstairs geometry each brane is accompanied by its image brane. Denoting by $[D_a]$ the homology class of the divisor $D_a$, we distinguish the three cases 
\begin{itemize}
  \item $[D_a] \neq [D_a']$,
  \item $[D_a] = [D_a']\qquad$ but $D_a \not= D_a'$ point-wise, and
  \item $D_a = D_a'\qquad\quad\,$ point-wise (D7-branes coincide with an O-plane)\;.
\end{itemize}
\vspace{0.2cm}

\noindent
In this article we are concerned with divisors of all three kinds and we would like to study the fate of the D-branes once we lift the orientifold model to a Calabi-Yau fourfold.

For stacks of D7-branes not invariant under the orientifold action the Chan-Paton gauge symmetry is $U(N_a)$, i.e.~it includes in particular the diagonal $U(1)_a\subset U(N_a)$. Each such stack of D7-branes can carry non-vanishing background flux for the Yang-Mills field strength $F_a$ supported on some two-cycles of $D_a$. All physical quantities depend only on the gauge invariant combination ${\cal F}_a = F_a + \iota^* B \mathbf{1}$ which involves the pullback of the B-field to the brane divisor. Only with non-vanishing gauge fluxes can one realize chiral spectra. In fact, the chiral index is simply given by
\beq
  I_{ab}= - \int_X [D_a ]\wedge [D_b]\wedge \big(\, c_1(L_a)-c_1(L_b)\, \big)
\eeq
in terms of the gauge flux $c_1(L_a) = \frac{1}{2 \pi} {\cal F}_a$. Note that in general this chiral matter is localised on the intersection of two divisors $D_a\cap D_b$, which defines a curve in $X$. The chiral spectrum can be further enhanced by vector-like pairs, which are detected by computing the relevant cohomology groups (for details we refer to the literature \cite{Blumenhagen:2008zz}).

In a global set-up the total charges of the orientifold planes and the D7-branes have to be cancelled. This includes the D7-brane tadpole cancellation condition
\beq \label{tadseven} 
  \sum_a  N_a\, ([D_a]+ [D'_a]) = 8\, [D_{\rm{O7}}], 
\eeq
where the sum is over all D7${}_a$-branes. 

The general condition for cancellation of the D3-brane tadpole is the most involved one and takes the form
\beq \label{D3_tadpolezwei} 
  N_{\rm{D3}} + \frac{N_{\rm{flux}}}{2} - \frac{1}{2} \sum_{a} \frac{1}{8\pi^2} \int_{D_a}\!\! {\rm tr}{\cal F}^2_a
  = \frac{N_{\rm O3}}{4} + \frac{\chi(D_{\rm O7})}{12} +\sum_{a} N_a\, \frac{\chi_{\mathrm{o}}(D_a)}{48},
\eeq
where the sum is understood over all branes $D_a$ and their image. Here $N_{\rm{D3}}$ counts the number of D3-branes and $N_{\rm flux}$ denotes the possible contributions from $G_3=F_3+\tau\, H_3$ form flux. The third term is due to the gauge flux background in the $U(N_a)$. The three contributions on the right hand side of \eqref{D3_tadpolezwei} are related to the O3-planes and the curvature induced terms on the
O7 and D7-branes.

The induced D3-charge on a smooth O7-plane is given by
\beq \label{eulerdiv}
  \chi(D_{\rm O7}) = \int_X \Big( [D_{\rm O7}]^3+ c_2(T_{X}) \wedge [D_{\rm O7}] \Big). 
\eeq
The contribution from the D7-branes is more involved as in Sen's orientifold limit \cite{Sen:1997gv}, since the D7-branes always intersect the O7-planes in double points \cite{Braun:2008ua, Collinucci:2008pf}. Therefore, the Euler characteristic is a priori not well defined. However, via the relation to F-theory it was argued in \cite{Collinucci:2008pf} that the correct Euler characteristic is
\beq \label{chi1}
  \chi_{\mathrm{o}}(D) = \int_X \Big( [D]^3 + c_2(X) \, [D] + 3 \, [D] \, [D_{\mathrm{O7}}] \, ([D_{\mathrm{O7}}]-[D]) \Big).
\eeq
The right-hand side of \eqref{D3_tadpolezwei} is precisely $\chi(Y)/24$ in the F-theory lift of this Type IIB orientifold, where $Y$ denotes the elliptically fibered Calabi-Yau fourfold. However, it is important to point out that in F-theory the fourfolds which yield non-Abelian gauge symmetries are not smooth.  To nevertheless compute the right-hand side of \eqref{D3_tadpolezwei} in the F-theory up-lift one has to resolve the singularities and determine the Euler characteristic of the smooth blow-up space.

\section{Orientifold geometries}
\label{secorientgeoms}
In \cite{Blumenhagen:2008zz,Grimm:2008ed} various compact Calabi-Yau geometries were considered which satisfy the two main conditions for realizing an $SU(5)$ GUT model:  
\begin{itemize}
  \item the Calabi-Yau $X$ contains shrinkable del Pezzo surfaces $D$
  \item there exist two-cycles on $D$ which are non-trivial in $D$ but trivial in $X$. 
\end{itemize}
As prototype examples and for concreteness, let us present two such simple geometries with their involutions. 

\subsubsection*{Single del Pezzo transition of the quintic}

The starting point is the familiar quintic $\IP^4[5]$, defined by a degree five hypersurface constraint in $\IP^4$ with homogeneous coordinates $x_i$, $i=1,\ldots 5$. At generic points in the complex structure moduli space the quintic $\IP^4[5]$ defines a smooth manifold. Choosing however the quintic polynomial as
\beq
  x_5^2\, P_3(x_1,x_2,x_3,x_4) + x_5 P_4(x_1,x_2,x_3,x_4)+ P_5(x_1,x_2,x_3,x_4)=0 ,
\eeq
with all monomials with factors $x_5^k,\, k>2$ vanishing, it degenerates such that at $(x_1,x_2,x_3,x_4,x_5)=(0,0,0,0,1)$ a del Pezzo singularity of the form $dP_6=\IP^{3}[3]$ is generated. From this singular locus one can deform to a new Calabi-Yau manifold, where this del Pezzo singularity is blown up to finite size and defines a new divisor of the Calabi-Yau manifold. These transitions can be described
via toric geometry and amount to introducing a new coordinate $x_6$ and a second projective equivalence.\footnote{See refs.~\cite{Blumenhagen:2008zz, Grimm:2008ed} for more details on these constructions.} The new degrees of the coordinates are shown below:

\begin{center}
  \begin{tabular}{r@{\,$=$\,(\,}r@{,\;\;}r@{,\;\;}r@{,\;\;}r@{\,)\;\;}|c|cc|c} 
    \multicolumn{5}{c|}{vertices of the} & coords & \multicolumn{2}{c|}{GLSM charges} & {divisor class}${}^\big.$ \\
    \multicolumn{5}{c|}{polyhedron / fan}      &        & $Q^1$   &            $Q^2$  &  \\ \hline\hline
    $v_1$ & $-1$ & $-1$ & $-1$ & $-1$ & $u_1$ & 1 & 0 & $H{}^\big.$ \\
    $v_2$ &   1  &   0  &   0  &   0  & $u_2$ & 1 & 0 & $H$  \\
    $v_3$ &   0  &   1  &   0  &   0  & $u_3$ & 1 & 0 & $H$  \\
    $v_4$ &   0  &   0  &   1  &   0  & $u_4$ & 1 & 0 & $H$  \\
    $v_5$ &   0  &   0  &   0  &   1  & $v$   & 1 & 1 & $H+X$  \\
    $v_6$ &   0  &   0  &   0  & $-1$ & $w$   & 0 & 1 & $X{}_\big.$  \\ \hline
    \multicolumn{5}{c|}{conditions:}  &       & 5 & 2 & ${}^\big.$ \\
  \end{tabular}
\end{center}

\noindent
In the following, the divisor $\{x=0\}$ is denoted as $D_x$. In this construction the divisors are first defined in the toric ambient space determined by the polyhedron. The Calabi-Yau hypersurface is then obtained as a representative of the anti-canonical class  
\beq \label{anti_class}
  \bar K = \sum_{i=1}^6 D_{x_i}  
\eeq 
of the toric ambient space. The toric divisors $D_x$ restrict to divisors of the hypersurface such that one can determine the triple intersections of this Calabi-Yau manifold by analyzing the intersections in the toric ambient space. By abuse of notation we will denote the divisors restricted to the hypersurface also by $D_x$. Introducing the basis $H=D_{u_1}$ and $X=D_{w}$ they read
\beq
  H^3 = 2, \qquad H^2 X = 3, \qquad H X^2 = -3, \qquad X^3 = 3 . 
\eeq
Note that these intersections become diagonal in the basis $X,\tilde H = H +X$ since 
\beq \label{diag_intdP6}
  \tilde H^3 = 5, \qquad  X^3 = 3 .
\eeq
reflecting the so-called swiss cheese property of this threefold. Using these intersections we can compute the Euler characteristics of the various divisors via \eqref{eulerdiv}. For instance for $D_w$ we consistently find $\chi(D_w)=9$. Note that this $dP_6$ divisor has seven non-trivial two-cycles, of which only one is non-trivial in the homology of the ambient Calabi-Yau threefold.
 
Next, one has to specify a holomorphic involution. Let us consider here the one acting as $\sigma: v\to -v$, which means that only even powers of this coordinate should appear in the constraint. Due to the two projective equivalences in the above table, the fixed point locus of this involution consists of the two disjoint smooth divisors  
\beq
  O7 = D_v + D_w
\eeq 
and no fixed points. The divisor $D_w$ is of course the $dP_6$ surface, while the divisor $D_v$ is smooth, non-rigid and has $\chi(D_v)=55$. 

Let us determine the  D7-tadpole canceling brane configuration with minimal gauge group. Naively, one might think that a single D7-brane wrapping a hypersurface in $[8\, H + 16\, X]$ might do the job. In this case, one would expect that the F-theory fourfold would only have $I_1$ singularities and the Weierstrass model is smooth, i.e.~the basis a Fano threefold. The D3-brane tadpole would then determine the Euler characteristic of the smooth fourfold via
\beq \label{naivchia}
  \chi^*(Y)=  \left( \frac{\chi_{\mathrm{o}}(8\, D_v)+ \chi_{\mathrm{o}}(8\,D_w)}{2}+ 2\chi(O7) \right)= 1728.
\eeq
However, in our case $D_w$ is rigid and as a consequence the best we can do is to cancel the induced O7-plane tadpole by a single brane along the divisor $8D_v$ and a stack of 8 D7-branes along the divisor $D_w$. The resulting gauge symmetry\footnote{Actually the gauge group is $SO(8)\times SO(1)$, where the second, trivial, factor is supported on the divisor $8D_v$.} is $SO(8)$ and for the Euler characteristic of the true uplifted fourfold $Y$ we find
\beq \label{1224}
  \chi(Y)= \left( \frac{\chi_{\mathrm{o}}(8\, D_v)+ 8\, \chi_{\mathrm{o}}(D_w)}{2} + 2\chi(O7) \right)= {1224}.
\eeq
Since the IIB model gives rise to a non-trivial non-Abelian gauge symmetry, the uplifted fourfold is expected to be generically \emph{singular} over the del Pezzo surface $D_w$. These singularities need to be resolved to compute the correct value of $\chi(Y)$. In the singular case the value $\chi^*(Y)$ is nevertheless of relevance. As we will make more precise below, $\chi^*(Y)$ is the Euler characteristic of the blown-up $Y$ plus the corrections due to the blow-up divisors.

\subsubsection*{Double del Pezzo transition of the quintic}

The second example is defined by one more del Pezzo transition and has two intersecting $dP_7$ surfaces. In this case the quintic polynomial is restricted such that all monomials containing $x_4^k,\, k>2$ or $x_5^m,\, m>2$ vanish. The resulting quintic now has two non-generic $dP_7=\mathbb{P}_{1,1,1,2}[4]$ singularities. Blowing these up into del Pezzo surfaces one introduces two additional coordinates $w_1$ and $w_2$ and two additional projective equivalences. The new scaling weights of the coordinates are shown below. 

\begin{center}
  \begin{tabular}{r@{\,$=$\,(\,}r@{,\;\;}r@{,\;\;}r@{,\;\;}r@{\,)\;\;}|c|ccc|c} 
    \multicolumn{5}{c|}{vertices of the} & coords & \multicolumn{3}{c|}{GLSM charges} & {divisor class}${}^\big.$ \\
    \multicolumn{5}{c|}{polyhedron / fan}      &        & $Q^1$   &
    $Q^2$ & $Q^3$  &  \\ \hline\hline
    $v_1$ & $-1$ & $-1$ & $-1$ & $-1$ & $u_1$ & 1 & 0 & 0 & $H{}^\big.$ \\
    $v_2$ &   1  &   0  &   0  &   0  & $u_2$ & 1 & 0 & 0 & $H$  \\
    $v_3$ &   0  &   1  &   0  &   0  & $u_3$ & 1 & 0 & 0 & $H$  \\
    $v_4$ &   0  &   0  &   1  &   0  & $v_1$ & 1 & 0 & 1 & $H+Y$  \\
    $v_5$ &   0  &   0  &   0  &   1  & $v_2$   & 1 & 1 & 0& $H+X$  \\
    $v_6$ &   0  &   0  &   0  &  $-1$  & $w_1$   & 0 & 1 & 0 & $X$  \\
    $v_7$ &   0  &   0  &  $-1$  & 0  & $w_2$   & 0 & 0& 1 & $Y{}_\big.$  \\ \hline
    \multicolumn{5}{c|}{conditions:}  &       & 5 & 2 & 2 & ${}^\big.$ \\
  \end{tabular}
\end{center}

In the basis $H=D_{u_1}$, $X=D_{w_1}$ and $Y=D_{w_2}$ the triple intersection numbers are 
\beq
  \bal
    & H^3 = 0, && H^2 X = 2, && H X^2 = -2, && X^3 = 2,\\
    & H^2 Y= 2, && H Y^2 = -2, && Y^3 = 2, && H X Y=1,\\
    & X^2 Y = -1, && X Y^2 = -1.
  \eal
\eeq
For the two $dP_7$ divisors $D_{w_1}$ and $D_{w_2}$  we consistently find $\chi(D_{w_i})=10$. Moreover, they intersect each other over a curve $C=\mathbb{P}^1$ of Euler characteristic
\beq \label{curve96}
  \chi(C)=-X\, Y\, (X+Y)=2\; .
\eeq
While there exist various holomorphic involutions, here we are interested in the one which exchanges the two $dP_7$ divisors:
\beq \label{involution_dP7}
  \sigma: \begin{cases}  v_1\leftrightarrow v_2, \\
                         w_1\leftrightarrow w_2. \end{cases} 
\eeq 
This implies that one element $D_- = X-Y$ is in $H_4^-(X)$, while the other combination $D_+ = X+Y$ is in $H_4^+(Y)$. In fact, consistent with \eqref{pmvanish}, one finds the intersections
\beq \label{dP7intersection2}
  \bal
    & D_+ H^2 = 4\ , && H D_+^2 = -2\ , && D_+^3 = -2\ , \\ 
    & D_+ D_-^2 = 6 \ ,&& H D_-^2 = -6\ .
  \eal
\eeq
As we will show one finds the intersections of elements in $H_4^+(X)$, the first line of \eqref{dP7intersection2}, in the base of the F-theory fourfold. 

Using the identifications $Q^2$ and $Q^3$ the fixed point locus of the involution \eqref{involution_dP7} is 
\beq \label{oplanecon}
  v_1\, w_1= v_2\, w_2 ,
\eeq
which defines a surface  in $[H+X+Y]$. Note that this is not the most generic surface in the homology class $[H+X+Y]$, as one could add a term $w_1\, w_2\, p_1({\bf u})$. Indeed looking at the common intersection of the O7-plane and the hypersurface constraint
\beq \label{hypersurf}
  \sum_{m,n=0}^2  v_1^m\, w_2^{2-m}\, v_2^n\, w_1^{2-n} p_{5-m-n}({\bf u})=0
\eeq
one finds a genus $g=0$ curve $\IP^1$ where in addition to \eqref{oplanecon} and \eqref{hypersurf} also $w_i=0$, and secondly a genus $g=6$ curve where also $v_i=0$.\footnote{The authors would like to thank the referee for pointing out a mistake here in the original version of the paper.} The $\IP^1$ curve is of course the same as the curve $C$ in eq.~\eqref{curve96} contained in both $dP_7$ divisors.

Now to cancel the tadpole one can introduce a single D7-brane wrapping a smooth surface in $[8(H+X+Y)]$. Since the O7-plane is smooth we can compute the Euler characteristic of the uplifted fourfold as 
\beq \label{naivedP7chi}
  \chi^*(Y)= \left( \frac{\chi_{\mathrm{o}}\big(\, 8\, (H+X+Y)\,\big)}{2} + 2\chi(H+X+Y) \right)= {1008},
\eeq
where we used $\chi(O7)=56$. Note that the fourfold $Y$ is truly singular. It is beyond the main scope of the present paper to compute these singular Euler characteristics. Note that in contrast to the first model, here in the uplifted fourfold we expect to find a generic singularity not over a divisor but only over a $\IP^1$ curve.



\section{F-theory uplifts}
\label{secuplifts}
Now we want to uplift these two orientifolds to F-theory on Calabi-Yau fourfolds. Recall that F-theory on an elliptically-fibered Calabi-Yau fourfold $Y$ with base $B$ is equivalent to Type IIB string theory on $B$ with a dilaton-axion $\tau=C_0 + \mathrm{i} e^{-\phi}$ varying over this base. In fact, at each point in $B$ the complex number $\tau$ can be identified with the complex structure modulus of the elliptic fiber over this point. If $Y$ is a Calabi-Yau manifold, the fiber degenerates over in general intersecting divisors $D_i$ in $B$ subject to the constraint
\beq \label{discr_a}
  \sum_i \delta_i \, D_i = 12 \, c_1(B).
\eeq
The $\delta_i$ denote the vanishing degree of the discriminant $\Delta$ over the divisor $D_i$ as listed for various enhancement types in appendix B. The relation (\ref{discr_a}) follows just from the fact that the descriminant is a section of $K_B^{-12}$. Hence, the powerful geometrical tools to analyze $Y$ allow to study string compactifications with strong coupling regimes.

To uplift the orientifold models we follow essentially the recipe of \cite{Collinucci:2008zs}. The idea is to first construct the base manifold $B$ and then to consider the Weierstrass fibration over this space. The base is given by the quotient $X/\sigma$ and can contain $\mathbb Z_2$ singularities related to the presence of O3-planes in the orientifold model. For the fourfold to be smooth the base of the fibration must be Fano, i.e. its anticanonical bundle $K_B^{-1 }$ must be ample. A criterion for this is that
\beq \label{Mori}
  -K_B \cdot C > 0
\eeq
for every effective curve. Since we are expecting the two uplifted fourfolds to be singular, the base manifolds will not be Fano.

\subsubsection*{Single del Pezzo transition}

Recall that the involution was $\sigma:v\to -v$, which led to two disjoint components for the O7-locus, namely $\{v=0\}$ and $\{w=0\}$. To describe the quotient $X/\sigma$ we are therefore looking for a map which is 2-to-1 away from the two O7-planes and 1-to-1 on them. This map can readily be defined as
\beq
  (u_1,u_2,u_3,u_4, v, w) \mapsto (u_1,u_2,u_3,u_4, v^2, w^2) .
\eeq
Now we consider the right-hand side as  new homogeneous coordinates and introduce $\tilde v=v^2$ and $\tilde w=w^2$. The toric data of the base threefold is shown below:

\begin{center}
  \begin{tabular}{r@{\,$=$\,(\,}r@{,\;\;}r@{,\;\;}r@{,\;\;}r@{\,)\;\;}|c|cc|c} 
    \multicolumn{5}{c|}{vertices of the} & coords & \multicolumn{2}{c|}{GLSM charges} & {divisor class}${}^\big.$ \\
    \multicolumn{5}{c|}{polyhedron / fan}      &        & $Q^1$   &            $Q^2$  &  \\ \hline\hline
    $v_1$ & $-1$ & $-1$ & $-1$ & $-2$ & $u_1$ & 1 & 0 & $P{}^\big.$ \\
    $v_2$ &   1  &   0  &   0  &   0  & $u_2$ & 1 & 0 & $P$  \\
    $v_3$ &   0  &   1  &   0  &   0  & $u_3$ & 1 & 0 & $P$  \\
    $v_4$ &   0  &   0  &   1  &   0  & $u_4$ & 1 & 0 & $P$  \\
    $v_5$ &   0  &   0  &   0  &   1  & $\tilde v$   & 2 & 1 & $2P+X$  \\
    $v_6$ &   0  &   0  &   0  & $-1$ & $\tilde w$   & 0 & 1 & $X{}_\big.$  \\ \hline
    \multicolumn{5}{c|}{conditions:}  &       & 5 & 1 & ${}^\big.$ \\
  \end{tabular}
\end{center}

\noindent
Let us discuss the construction of the hypersurface which is the base $B$. To begin with, we proceed as in the Calabi-Yau case and find the maximal triangulations of the ambient toric space obtained from the polyhedron. However, we then do not consider the hypersurface representing the anti-canonical class $\bar K = \sum_i D_i$ of the ambient toric space as in \eqref{anti_class}, but rather the one corresponding to
\beq
  \bar K - X - P  .
\eeq
Note that this is not a Calabi-Yau manifold as $c_1(B)=P+X$. The Stanley-Reisner ideal of the ambient toric space is $\{u_1 u_2 u_3 u_4 =0\}\cup\{\tilde v\tilde w=0\}$ such that the resulting triple intersection form on $B$ reads
\beq
  P^3 = 1, \qquad P^2 X = 3, \qquad P X^2 = -6, \qquad X^3 = 12 . 
\eeq
This is readily compared to the intersection form  \eqref{diag_intdP6} in the orientifold model if one takes the basis $X,\tilde P = 2P + X$ where
\beq
  \tilde P^3 = 20 ,\qquad X^3 = 12 .
\eeq
As expected the two O7-components do not intersect. For the Euler characteristics of the base divisors we find
\beq
  \chi(P)=18, \qquad \chi(2P+X)=55 ,\qquad  \chi(X)=9 ,
\eeq
which shows that the O7-planes do not change their topology.

The fourfold $Y$ is now defined as the Weierstrass model over this base,
\beq \label{weierstrass}
  y^2=x^3 + x\, z^4\, f(u_i,\tilde v,\tilde w) + z^6\, g(u_i,\tilde v,\tilde w),
\eeq
where $z$ is the section of the fibration, i.e.~the divisor $Z=\{z=0\}$ is the base $B_3$. For the hypersurface constraint \eqref{weierstrass} to be well defined $f$ and $g$ have to be sections of appropriate line bundles, i.e.~$f\in H^0(B;K^{-4}_B)$ and $g\in H^0(B;K^{-6}_B)$ with the canonical line bundle given by $K_B={\cal O}(-P-X)$. Note that the total Calabi-Yau fourfold is a complete intersection of two hypersurfaces. In fact, the base $B$ is not of Fano type, as can be seen by checking the criterion (\ref{Mori}) for the effective curve $C= D_{\tilde w} \cap  D_{u_1}$,
\beq
  -K_B \cdot X\cdot P = -3 .
\eeq

Since the fourfold is singular, we cannot directly compute $\chi(Y)$. For such a computation one would first have to resolve the singularities to obtain a smooth $Y$. Recall that only in the case of a smooth fourfold with only $\mathrm{I}_1$ singularities of the elliptic fibration does one have \cite{Sethi:1996es, Klemm:1996ts}
\beq \label{klemmyau} 
  \chi^*(Y)=  12 \int_B c_1(B)\, c_2(B) + 360 \int_B c^3_1(B)\ ,
\eeq
where $B$ is the base of the fibration. If we naively apply this equation to the non-Fano base $B$ just constructed, we find $\chi^*(Y)=1728$. This was also the value \eqref{naivchia} for the naive cancellation of the tadpoles in the orientifold model. This can be understood as follows. The correct Euler characteristic of the blown-up $Y$ is $\chi^*(Y)-\delta$, where $\delta$ is a correction term which depends on the Chern classes of the divisors, curves and points over which the fiber degenerates. For example, if the fiber only degenerates over a divisor $D$ with gauge group $G$ the corrected Euler characteristic of the blowup space is given by \cite{Andreas:1999ng}
\beq
  \chi(Y)=  \chi^*(Y) - r_{G}\, c_G \, ( c_G +1) \int_D c_1(D)^2 ,  
\eeq
where $r_G$ and $c_G$ are the rank and dual Coxeter number of $G$. This can be generalized to more complicated degenerations in higher codimension~\cite{Andreas:1999ng, Andreas:2009uf}. In our example, for an $SO(8)$ singularity along $D_{\tilde w}$ (and no additional non-abelian enhancement  over any other divisor on $B$), generically no such higher codimension degenerations occur, and the above formula correctly reproduces $\chi(Y)=1224$ as computed in the orientifold picture in (\ref{1224}). Roughly speaking $\chi^*(Y)$ can be understood as the leading contribution, which will then receive corrections due to the singularities. On the one hand, in the orientifold picture this correction cannot be switched off due to the rigidity of D7-branes on the del Pezzo surface. On the other hand, in the F-theory fourfold there exist no complex structure deformations which allow one to remove the gauge enhancement while preserving the Weierstrass form. Clearly, this matches the fact that deformations of D7-branes are mapped to complex structure deformations of the Calabi-Yau fourfold. Nevertheless, if we subtract these corrections on both sides, the matching of the easily computable numbers $\chi^*(Y)$ with base $B$ still provides a non-trivial check of the F-theory up-lift of orientifold models.

In section \ref{sec:tate} we will discuss what kind of degenerations and gauge theory enhancements can arise in this singular Weierstrass model and compare it to the orientifold expectation. But first let us present the uplift of the exchange orientifold.

\subsubsection*{Double del Pezzo transition}

For this orientifold we proceed in exactly an analogous manner. The required map which is 1-to-1 on the orientifold locus $v_1\, w_1= v_2\, w_2$ and 2-to-1 away from it is
\beq
  (u_1,u_2,u_3, v_1, v_2, w_1, w_2) \mapsto (u_1,u_2,u_3,v_1\,v_2, w_1 w_2, v_1 w_1 + v_2 w_2) .
\eeq
Note that the right-hand side has one coordinate less than the left-hand side, which is expected as the holomorphic involution identifies two coordinates. Similarly, we also expect that one projective equivalence drops out. Indeed after introducing new homogeneous coordinates $v=v_1 v_2$, $w=w_1 w_2$, $h=v_1 w_1+v_2 w_2$, the base manifold is described by the toric data below: 

\begin{center}
  \begin{tabular}{r@{\,$=$\,(\,}r@{,\;\;}r@{,\;\;}r@{,\;\;}r@{\,)\;\;}|c|cc|c} 
    \multicolumn{5}{c|}{vertices of the} & coords & \multicolumn{2}{c|}{GLSM charges} & {divisor class}${}^\big.$ \\
    \multicolumn{5}{c|}{polyhedron / fan}      &        & $Q^1$   &            $Q^2$  &  \\ \hline\hline
    $v_1$ & $-1$ & $-1$ & $-2$ & $-1$ & $u_1$ & 1 & 0 & $P{}^\big.$ \\
    $v_2$ &   1  &   0  &   0  &   0  & $u_2$ & 1 & 0 & $P$  \\
    $v_3$ &   0  &   1  &   0  &   0  & $u_3$ & 1 & 0 & $P$  \\
    $v_4$ &   0  &   0  &   1  &   0  & $v$ & 2 & 1 & $2P+X$  \\
    $v_5$ &   0  &   0  &   0  &   1  & $h$   & 1 & 1 & $P+X$  \\
    $v_6$ &   0  &   0  & $-1$  & $-1$ & $w$  & 0 & 1 & $X{}_\big.$  \\ \hline
    \multicolumn{5}{c|}{conditions:}  &       & 5 & 2 & ${}^\big.$ \\
  \end{tabular}
\end{center}

The resulting triple intersection form on $B$ reads
\beq \label{dP7base}
  P^2 X = 2, \qquad P X^2 = -1, \qquad X^3 = -1,
\eeq
leading to
\beq
  \chi(P)=13, \quad \chi(2P+X)=46,\quad  \chi(X)=10,\quad \chi(2P+2X)=56 .
\eeq
Therefore, $X$ can be identified as the invariant $dP_7$ divisor $D_+$ introduced before \eqref{dP7intersection2} and $2P+2X$ as the divisor class of the orientifold plane. Indeed, as expected the triple intersections \eqref{dP7base} for the fourfold base and the first line in the orientifold intersections \eqref{dP7intersection2} agree up to an overall factor $2$.

The fourfold $Y$ is again defined as the Weierstrass model over this base
\beq \label{weierstrass2}
  y^2=x^3 + x\, z^4\, f(u_i,v,h, w) + z^6\, g(u_i,v,h, w),
\eeq
giving again a (singular) complete intersection of two hypersurfaces. As anticipated, the base $B$ is not of Fano type as can seen by computing for the effective curve $C= D_{w} \cap  D_{h}$
\beq
  -K_B \cdot X\cdot (P+X) = -1 .
\eeq
Therefore a smooth $Y$ to determine $\chi(Y)$ has to be obtained by blowing up the singularities. However, as above, we can employ \eqref{klemmyau} to compute the leading order Euler characteristic $\chi^*(Y)=1008$ which has to be corrected by the topological data of the blow-ups. Again this matches with the naive computation for the orientifold side \eqref{naivedP7chi}.

\section{Consequences for F-theory GUTs}
\label{sec:tate}

Since in the last two sections we have established the F-theory lift of certain orientifold models, we can now study and compare the possible gauge theory enhancements. In the Type IIB orientifold construction these are given by the Chan-Paton factors for the D7-branes. As is well known this gives only rise to adjoint, bifundamental, symmetric and antisymmetric representations of $U(N)$, $SO(N)$ and $SP(N)$. Clearly, this excludes all exceptional gauge groups. On the F-theory side the gauge group is encoded in the degenerations of the elliptic fiber over the base manifold. Since we have the explicit Weierstrass model available, we can use the Tate algorithm to determine the possible types of degenerations. In principle we could imagine classifying all possible gauge groups that arise in this way. However, in this paper we will only be discussing certain interesting cases. Moreover, in this article we do not consider any $G_4$-form flux, which would be the uplift of gauge fluxes on the D7-branes. One should however keep in mind that in a fully consistent model the Freed-Witten anomaly forces us to have some fluxes non-vanishing.\footnote{For examples of this phenomenon in the IIB limit see \cite{Blumenhagen:2008zz}.}

\subsubsection*{Tate algorithm}

For completeness we present here a short explanation of how the Tate algorithm is used to determine the degeneration of the elliptic curve \cite{Bershadsky:1996nh}. Instead of the reduced form \eqref{weierstrass}, one uses the complete Tate form of the hypersurface constraint
\beq
  y^2 + x\, y\, z\, a_1+ y\, z^3\, a_3= x^3 + x^2\, z^2 \, a_2 + x\, z^4\, a_4 + z^6\, a_6 ,
\eeq
where the $a_n$ are sections of $K_B^{-n}$. In terms of the combinations
\beq
  b_2=a_1^2 + 4\, a_2, \qquad b_4=a_1\, a_3 +2 a_4, \qquad b_6=a_3^2+4\, a_6
\eeq
the functions $f$ and $g$ in the Weierstrass form are given by
\beq
  f=-\frac{1}{48}( b_2^2 -24\, b_4), \qquad
  g=-\frac{1}{864}( -b_2^3 + 36 b_2 b_4 -216 \, b_6) .
\eeq
The discriminant can then be expressed as 
\beq \label{DeltaF}
  \Delta_F =-{\textstyle \frac{1}{4}}\, b_2^2\, (b_2 b_6- b_4^2) - 8 b_4^3 -27 b_6^2 +9 b_2 b_4 b_6 .
\eeq
As detailed in \cite{Bershadsky:1996nh}, the possible singularities of the fiber and the corresponding gauge groups can be read off almost entirely from the vanishing order of the sections $a_n$ and of the discriminant on the location of the D7-branes. For convenience of the reader we have collected the required data in appendix \ref{appb}. We recall from  \cite{Bershadsky:1996nh} that a vanishing degree $d$ of $a_n$ in the table along divisor $D_x= [x=0]$ is to be interpreted in the sense that $a_n$ can be written as $a_{n,d} \, x^d$.

To recover Sen's orientifold limit \cite{Sen:1997gv} one rescales the $b_i$ as\footnote{Equivalently, one can rescale $a_3\to \epsilon a_3$, $a_4\to \epsilon a_4$, $a_6\to \epsilon^2 a_6$ and send $\epsilon\to 0$.}
\beq
  b_2= - 12 \, h, \qquad b_4 = 2 \, \epsilon \, \eta , \qquad b_6= - \frac{\epsilon^2}{4} \chi.
\eeq
The orientifold limit is defined by taking $\epsilon \rightarrow 0$ such that the string coupling becomes weak away from $h=0$. The leading order discriminant then takes the form
\beq \label{Deltae}
  \Delta_\epsilon = - 9 \epsilon^2 h^2 (\eta^2 - h \chi)  \ 
\eeq
plus corrections cubic or higher in $\epsilon$. This is just the first term in the full F-theory discriminant \eqref{DeltaF}. The D7-branes and O7-planes are thus located at
\beq \label{O7D7}
  \bal
    &\text{O7:} && h = 0  , \\
    &\text{D7:} && \eta^2 = h \chi.
  \eal
\eeq
The Type IIB theory is defined on the Calabi-Yau threefold $X$ which is a double cover of the base $B$ branched over $h=0$. Note that (\ref{O7D7}) shows already that the defining equation for a D7-brane configuration with a well-defined F-theory uplift is non-generic \cite{Braun:2008ua,Collinucci:2008pf}.

Away from the limit $\epsilon \rightarrow 0$ the factorization of the discriminant into a perturbative O7-plane and the perturbative D7-branes is generically lost. The F-theoretic description takes into account the non-perturbative effects which smoothen out the singular orientifold plane. For a single stack of D7-branes this does not mean that the gauge symmetry on the D7-branes is automatically reduced to lower rank in the F-theory. Rather, as long as the singularity type of the higher order terms in $\epsilon$ is worse in the sense of Tate's algorithm, the perturbative gauge group persists also in the full F-theory. More generally, unless  $\Delta_F$ exhibits the same factorization properties as $\Delta_\epsilon$ the rank of the (product) gauge group can be reduced along a non-perturbative Higgs branch.

\subsection{Single del Pezzo transition}

Let us now consider a couple of interesting degenerations of the elliptic fiber occurring in the uplift of the first orientifold model.

\subsubsection*{Non-Abelian gauge symmetry on the orientifold locus}

We first consider the sublocus in complex structure moduli space with a well-defined orientifold limit. Recall that $\{b_2=a_1^2+4 \, a_2=0\}$ defines the location of the O7-planes. Taking into account that $a_n\in H^0\big(B; {\cal O}(\, n(P+X)\,)\big)$, the general form of $a_1$ and $a_2$ is 
\beq \label{oplanetat}
  a_1=p_1({\bf u}) \, \tilde w, \qquad 
  a_2= c_0\, \tilde v\, \tilde w + p_2({\bf u})\, \tilde w^2.
\eeq
Here $p_n({\bf u})$ denotes a polynomial of degree $n$ in $u_1,u_2,u_3$ and $c_0\in \mathbb C$ a complex structure modulus. Requiring that the F-theory really describes the uplift of the orientifold fixes some of the complex structure moduli in \eqref{oplanetat} such that the O7-plane is located at $\tilde v \, \tilde w =0$. This translates into $p_2({\bf u})=-\frac{1}{4}p^2_1({\bf u})$.

The simplest brane configuration in Type IIB occurs when we cancel the orientifold charge locally by placing a stack of 8 branes on top of the divisor $D_v$ and a stack of $8$ branes on top of $D_w$. Without any gauge flux the resulting gauge group is simply $SO(8) \times SO(8)$. The expected Euler characteristic of the singular fourfold is
\beq
  \chi(Y)=\left(  \frac{8 \chi_{\mathrm{o}}(H+X) + 8 \chi_{\mathrm{o}}(X)}{2} + 2 \chi(O7) \right) = 384.
\eeq
Consistently, this number reduces considerably by introducing stacks of D7-branes. In F-theory this effect must be captured by correctly defining $\chi(Y)$ for the now singular fourfolds \cite{Klemm:1996ts,Andreas:1999ng,Andreas:2009uf}. 

For the special case of local orientifold charge cancellation the coupling constant $g_s$ is a constant everywhere and can be taken as a free parameter. As a first consistency check, the corresponding gauge group must also be reproduced from the F-theory point of view. Indeed application of Tate's algorithm with the help of appendix \ref{appb} identifies the corresponding F-theory configuration with gauge group $SO(8) \times SO(8)$ as
\beq \label{tateSen1}
  a_1=0, \quad  
  a_2= \tilde v\, \tilde w, \quad  
  a_3 =0, \quad 
  a_4 = c_1 ( \tilde v \, \tilde w)^2, \quad 
  a_6=0. 
\eeq

Next we determine the possible maximal non-Abelian gauge symmetry in the orientifold limit both from a IIB and from an F-theoretic point of view. In the orientifold model, one can cancel the D7-brane tadpole by introducing a stack of eight D7-branes wrapping the divisor $D_{u_1}=H$ and sixteen D7-branes wrapping the $dP_6$ surface $D_{w}=X$. Without any gauge flux, this yields gauge group $SP(8)\times SO(16)$. The expected Euler characteristic of the singular fourfold is
\beq
  \chi(Y)=\left(  \frac{8 \chi_{\mathrm{o}}(H) + 16 \chi_{\mathrm{o}}(X)}{2} + 2 \chi(O7) \right) = 312.
\eeq
Since $H$ and $X$ intersect there exists also non-chiral matter on the intersection curve. The question now is whether one can find this gauge group also in the F-theory lift, i.e.~whether one can arrange for the elliptic fiber to degenerate such as to produce an $SP(8)$ singularity over $D_{u_1}$ and an $SO(16)$ over $D_{\tilde w}$. A priori it is not excluded that this gauge symmetry is reduced once effects non-perturbative in $g_s$ are taken into account. The point is that for configurations where the charge of the orientifold plane is not cancelled locally, $g_s$ cannot be considered as constant or taken to be arbitrarily small everywhere. While in F-theory the backreaction of the D-branes is fully taken into account, the orientifold approach treats all branes in the probe approximation, and extra surprises might happen.

Instead one finds that it is still possible to achieve this maximal enhancement along $u_1=0$ and $\tilde w=0$ on the sublocus in complex structure moduli space where
\beq \label{tatemax}
  \bal
    & a_1= p_1({\bf u}) \, \tilde w, &&\quad  a_2=\tilde v\, \tilde w - {\textstyle \frac{1}{4}} \big(p_1({\bf u}) \, \tilde w\big)^2, &&\quad a_3= 0, \\
    & a_4= c_1 \, u_1^4 \tilde w^4,  &&\quad  a_6=0 .
  \eal
\eeq
On this locus the non-Abelian gauge group on $D_{u_1}$ is $SP(8)$ and the gauge group on $D_{\tilde w}$ is $SO(16)$.

On the other hand, non-perturbative effects still leave their imprint on the geometry. To see this we compare the expressions (\ref{DeltaF}) and (\ref{Deltae}) for the discriminant locus in Sen's limit and in the full F-theory,
\beq
  \Delta_{\epsilon} = 16 c_1^2\, u_1^8 \, \tilde w^{10} \, \tilde v^2,  \quad \quad 
  \Delta_{F} = 16 c_1^2\, u_1^8 \, \tilde w^{10} \, (\tilde v^2 - 4 c_1 u_1^4 \tilde w^2).
\eeq
What happens is that once the genuinely F-theoretic higher corrections in $\Delta$ are taken into account, the component $\tilde v =0$ of the O-plane, whose charge is not cancelled locally, splits into two objects $\tilde v = \pm 2 \sqrt{c} u_1^2 \tilde w$. This splitting of the O-plane into two non-perturbative 7-branes in F-theory is familiar from compactifications to eight \cite{Sen:1996vd} and six \cite{Sen:1997kw,Sen:1997gv} dimensions. We find it interesting to have an explicit laboratory to study this effect in four-dimensional vacua.

As elaborated in section \ref{secorientgeoms}, we are expecting that there also exists a minimal non-Abelian gauge symmetry on the $dP_6$ surface. Indeed it is easy to see that the sections $a_n$ cannot avoid some overall factors in $\tilde w$, i.e. 
\beq
  a_n=\tilde w^{d_n}(\ldots )\qquad  {\rm with}\ (d_1,d_2,d_3,d_4,d_6)=(1,1,2,2,3)
\eeq
which according to the Tate algorithm gives a $G_2$ singularity. In the Sen limit this exceptional gauge group gets exhanced to $SO(8)$, as $a_6/a_i\to 0$ for $\epsilon\to 0$ and $i=3,4$ \footnote{Note that due to the quadratic and, respectively, linear dependence of $b_6$ on $a_3$ and $a_6$ we do not neglect $a_6$ in the (perturbative) discriminant but only in determining the maximal vanishing degrees of the $a_i$ in the table in appendix B, which are now $(1,1,2,2,4)$.}. Thus, the F-theory minimal gauge group is smaller than the perturbative one.

\subsubsection*{Exceptional gauge groups} 

From eqs.~\eqref{oplanetat} and \eqref{tatemax} and table \ref{tab:TateTable} it is clear that in the orientifold uplift we can only get orthogonal gauge groups on the del Pezzo surface $D_w$. However, by choosing the complex structure such that $c_0=0$ in \eqref{oplanetat}, we have a chance to also find exceptional gauge groups. In this case $b_2\ne \tilde v \tilde w$ and this does not correspond to any orientifold model. Let us exemplify this with an $E_6$ singularity over the $dP_6$ surface. This can be engineered by choosing
\beq \label{tatemax2}
  \bal
    & a_1=p_{(1,0)}\, \tilde w,   &&\quad a_2= p_{(2,0)}\, \tilde w^2, &&\quad a_3= p_{(3,1)}  \tilde w^2, \\
    & a_4=p_{(4,1)}\, \tilde w^3, &&\quad a_6= p_{(6,1)}\, \tilde w^5,
  \eal
\eeq
with $p_{(m,n)}$ denoting a section of ${\cal O}(m P + n X)$ which is not just of the form $p_4({\bf u}) \tilde w$. For this choice one gets an $E_6$ singularity, which is enhanced to $E_7$ for $p_{(3,1)}=0$. On the curve $\tilde w=p_{(3,1)}=0$ we thus find matter fields in the fundamental ${\bf 27}$ representation. The singularity is further enhanced to $E_8$ where in addition $p_{(4,1)}=0$. Therefore, on the intersection locus
\beq
  \{w=p_{(3,1)}=p_{(4,1)}=0\}
\eeq
one finds the Yukawa couplings ${\bf 27}^3$. The number of these points is $X\, (3P+X)\, (4P+X)=6$. We conclude that on the uplifted orientifold base, it is possible to engineer exceptional gauge symmetries by moving away from the orientifold locus in F-theory complex structure moduli space.

\subsubsection*{Spinors of {\itshape SO}(10)} 

Having realized $E_6$ it is then natural to ask whether one can also start with $SO(10)$ on the del Pezzo and find the spinor representation on some curve where the fiber is enhanced to $E_6$. The sections for realizing $SO(10)$ on $D_w$ are
\beq \label{tateso10-a}
  \bal
    & a_1=p_{(1,0)}\, \tilde w,  &&\quad  a_2= p_{(2,1)}\, \tilde w, &&\quad a_3= p_{(3,1)}\,  \tilde w^2, \\
    & a_4=p_{(4,1)}\, \tilde w^3,  &&\quad  a_6= p_{(6,1)}\, \tilde w^5  ,
  \eal
\eeq
which is enhanced to $E_6$ on the curve $w=p_{(2,1)}=0$. However, for our base we have $X(2P+X)=0$ so that the intersection is empty. This can be traced back to the fact that the two orientifold planes in the Calabi-Yau do not intersect. We conclude that while it is possible with the uplift of our first orientifold to get exceptional gauge symmetries, spinors of $SO(10)$ are still not possible. What one could do however, is to break the $E_6$ to $SO(10)$ via additional $U(1)$ fluxes. In this way, one would also get spinors of $SO(10)$ and the ${\bf 16}\,{\bf 16}\, {\bf 10}$ Yukawa coupling but at the price of introducing potential exotic matter states.

\subsection{Double del Pezzo transition}

Let us now consider the uplift of the second orientifold defined via the double $dP_7$ transition with the exchange orientifold projection. In this case $a_1$ and $a_2$ can have more terms
\beq \label{oplanetat2}
  a_1=c_h\, h+ p_1({\bf u})\, w, \qquad 
  a_2= c_0\,  v\, w + c_{h^2}\, h^2 + q_1({\bf u})\, h\, w+ p_2({\bf u})\, w^2.
\eeq
For the orientifold uplift, we expect $b_2$ to be given by
\beq
  b_2=\eta ( h^2 - 4 v\, w),
\eeq
where $\eta\ne 0$ is some unknown constant. This restricts $a_2$ to take the form
\beq
  a_2= -\eta\,  v\, w  +\frac{\eta-c_{h^2}}{4}\, h^2 -\frac{c_h}{2} p_1({\bf u})\, h\, w- \frac{1}{4} p^2_1({\bf u})\, w^2 .
\eeq
Recall from our discussion around eq.~\eqref{hypersurf} that in this case, we are not expecting a generic degeneration of the Weierstrass fibration over a surface but only over a rigid curve $\IP^1$. A closer look reveals that generically the  sections $a_n$ cannot avoid some overall factors in $g=w=h$, i.e. 
\beq
  a_n = g^{d_n}(\ldots )\qquad  {\rm with}\ (d_1,d_2,d_3,d_4,d_6)=(1,1,2,2,3),
\eeq
which gives a $G_2$ singularity over the genus zero curve $D_{w}\cap D_{h}$. 

On the complex structure locus $c_h=\eta\ne 0$ along the $dP_7$ divisor $D_w=X$, $SU(N)$ degenerations are possible. This agrees with our expectations from the orientifold, as in the upstairs Calabi-Yau threefold the two $dP_7$s get exchanged and thus carry unitary Chan-Paton labels. The maximal gauge group in Type IIB theory is now  $SP(8) \times SU(8)$, corresponding to the D7-tadpole canceling configuration of 8 branes on, say, $D_{u_1}$ and $D_{w_1}$ (plus their image on $D_{w_2}$).  This configuration is easily matched in F-theory for
\beq \label{tateso10}
  \bal
    & a_1= c_h h + p_1({\bf u}) w,    &&\quad  a_2=-c_h \, v \, w - \frac{c_h}{2}  \,  p_1({\bf u})  \, h\, w -\frac14 p^2_1({\bf u}) w^2,  &&\quad a_3= 0, \\
    & a_4= c_d \, u_1^4 w^4, &&\quad  a_6= 0.
  \eal
\eeq
A similar splitting of the O7-plane is observed as for the previous model.

Let us study possible GUT enhancements off the orientifold locus. By choosing $c_h=c_{h^2}=0$ in \eqref{oplanetat2}, we can also arrange for an orthogonal gauge group along $D_w$. For sections $a_i$ as in \eqref{tateso10-a} one finds an $SO(10)$ singularity with the potential spinors supported on $w=p_{(2,1)}=0$. However, for the second base $X(2P+X)\ne 0$ but instead defines a genus $g=1$ curve on $dP_7$. The Higgs fields in the ${\bf 10}$ representation are localized on the genus $g=4$ curve $w=p_{(3,1)}=0$ and the ${\bf 16}\,{\bf 16}\, {\bf 10}$ Yukawa coupling on the points $w=p_{(2,1)}=p_{(3,1)}=0$. There are $X(2P+X)(3P+X)=6$ such points.

Therefore, for the exchange involution, after moving away from the orientifold locus, we can get $SO(10)$ GUTs with spinor and vector representations and the ${\bf 16}\,{\bf 16}\, {\bf 10}$ Yukawa couplings\footnote{This seems to be in contrast to the statement made in \cite{Donagi:2009ra} that for uplifted orientifold models the spinor representation of $SO(10)$ would be  localized on a curve of $\mathbb Z_2$ singularities. However, there the assumption was made that the threefold is of  the form $\xi^2=b_2$ with involution $\xi\to-\xi$, which is not the case for our examples.}.

\subsubsection*{Yukawas ${\bf 10}\,{\bf 10}\,{\bf 5}_{{\bf H}}$ of {\itshape SU}(5)} 

We conclude by studying whether for the second fourfold one can also find $SU(5)$ GUTs with ${\bf 10}$ and ${\bf \ov 5}$ representation and in particular the perturbatively absent ${\bf 10}\,{\bf 10}\,{\bf 5}_{{\bf H}}$ Yukawa couplings. From the Tate algorithm, the sections for realizing $SU(5)$ on $D_w$ are 
\beq \label{tatesu5}
  \bal
    & a_1=p_{(1,1)}, \,   &&\quad  a_2= p_{(2,1)}\, w, &&\quad a_3= p_{(3,1)}\,  w^2, \\
    & a_4=p_{(4,1)}\,  w^3,  &&\quad  a_6= p_{(6,1)}\, w^5.
  \eal
\eeq
This gets enhanced to $SO(10)$ along the genus $g=0$ curve
\beq
  SO(10): \quad \{w=p_{(1,1)}=0 \}
\eeq
supporting matter in the ${\bf 10}$ representation. In addition one finds matter in the ${\bf 5}$ representation on the curve of $SU(6)$ enhancement
\beq
  SU(6): \quad \{w= p_{(3,1)}^2 p_{(2,1)} - p_{(4,1)} p_{(3,1)} p_{(1,1)} + p_{(6,1)} p_{(1,1)}^2 = 0\}
\eeq
as follows from an explicit analysis of the discriminant. Extra enhancement to $SO(12)$ occurs for $\{w=p_{(1,1)}=p_{(3,1)}=0 \}$. At this $X(P+X)(3P+X)=1$ point the bottom-quark Yukawa couplings ${\bf 10}\,{\bf \ov 5}\, {\bf \ov 5}_{{\bf H}}$ are localized. The ${\bf 10}\,{\bf 10}\, {\bf 5}_{{\bf H}}$ Yukawa couplings would be localized at the further $E_6$ enhancement $\{w=p_{(1,1)}=p_{(2,1)}=0 \}$. However, on our second base $X(P+X)(2P+X)=0$ so that this  F-theory model fails to support the top-quark Yukawas.

\section{Conclusions}

In this note we have found the F-theory uplift of  Calabi-Yau threefolds which contain shrinkable del Pezzo surfaces with non-trivial relative homology. Such geometries had been used in \cite{Blumenhagen:2008zz} to implement phenomenologically appealing $SU(5)$ GUT models into Type IIB orientifold compactifications. Our method, inspired by \cite{Collinucci:2008zs}, has been to first define a
base threefold as the orientifold quotient of the Calabi-Yau threefold and then to consider a Weierstrass model thereof. The resulting Calabi-Yau fourfold can be described as a complete intersection of a toric ambient sixfold. We have seen that orientifolds with stacks of D7-branes on del Pezzo surfaces generically lead to singular fourfolds with a degenerate Weierstrass fibration over a non-Fano base. These singularities have to be resolved in order to compute the fourfold Euler characteristic and match the geometric D3-brane tadpole in the orientifold models.

We have analyzed explicitly which special subloci of the complex structure moduli space of these fourfolds correspond in Sen's limit to a perturbative IIB compactification. With the help of Tate's algorithm we have engineered some possible gauge groups arising on the degenerations of the fourfold both for models with an orientifold limit and for more general configurations. As one of our main findings we have identified smooth deformations in the F-theory complex structure moduli space which take a configuration with a perturbative IIB limit to a setup with only non-perturbatively possible exceptional gauge groups or spinor representations. As far as possible applications to GUT model building are concerned we have realized an $SO(10)$ GUT group with spinor representations and the necessary structure for Yukawa couplings. The top Yukawa couplings for a GUT $SU(5)$ could however not be obtained from the specific uplifts we consider here. It will be interesting to see whether more general complete intersection Calabi-Yau fourfolds  can  permit them.


\subsection*{Acknowledgements}

We gratefully acknowledge discussions with T.~W.~Ha, H.~Jockers, D.~Klevers, E.~Plauschinn and especially A.~Klemm. 
RB~would like to thank the Galileo Galilei Institute for Theoretical Physics for hospitality and the INFN for partial support during the completion of this work. 
TG~and TW thank the KITP Santa Barbara and the MPI Munich for hospitality.
The computations of the toric geometry where carried out using \cite{Rambau:TOPCOM-ICMS:2002} and the Maple package {\tt schubert}.
TW is supported by the DOE under contract DE-AC03-76SF00515.
This work was supported in parts by the European Union 6th framework program 
MRTN-CT-2004-503069 ``Quest for unification'', 
MRTN-CT-2004-005104 ``ForcesUniverse'', 
MRTN-CT-2006-035863 ``UniverseNet'',
SFB-Transregio 33 ``The Dark Universe'' by the DFG.


\appendix

\newcommand{\KK}{\mathrm{K3}}   
\newcommand{\Kb}{\KK^*}         

\section{More F-theory uplifts}
\label{appa}
In addition to the examples covered in sections \ref{secorientgeoms} and \ref{secuplifts}, the outlined uplifting procedure was applied to a number of further orientifold involutions and base geometries. 

The data is read as follows: The left column contains the orientifold information, where the 'base space' refers to the upstairs geometry of the threefold $X$. The Euler characteristic is found under 'topology'. Coordinates of the del Pezzo-transitions of the quintic (tables \ref{tab:upliftsdP6} and \ref{tab:upliftsdP7sq}) are in accordance to the main text. For the conventions of the later examples of elliptically-fibered threefolds over del Pezzo bases we refer to section 4 of \cite{Blumenhagen:2008zz}. The naming scheme $\smash{M_n^{(dP_{i_1},\dots,dP_{i_n})}}$ refers to the threefold with base $dP_n$, where the divisors of the last $n$ coordinates are of type $dP_{i_1},\dots,dP_{i_n}$ in the respective order. Note that $dP_9$ and $dP_{10}$ are not del Pezzo surfaces, but are likewise defined by blowing up $\IP^2$ at 9 or 10 distict points.

In the right column the corresponding uplift data of the elliptically-fibered Calabi-Yau fourfold is provided. The added divisor corresponds to the coordinate of weight 1 in the $\IP_{3,2,1}[6]$ elliptic fiber and embeds the downstairs threefold base $B$ into the fourfold. Under 'topology' one finds the Euler characteristic of the downstairs base and the O7-planes with respect to the new coordinates. Finally, we compute the Euler characteristic of the fourfold, where we find perfect agreement to the predicted value from the threefold side in each case. The symbol $\chi^*$ for the fourfold Euler characteristic refers to the fact that $Y$ might be singular, refer to section \ref{secuplifts} for a proper definition.

\begin{table}[ht]
        \centering
                \begin{tabular}{ll|ll}
                  \multicolumn{2}{c|}{orientifold data (3-fold)} & \multicolumn{2}{c}{uplift data (4-fold)${}_\big.$} \\ \hline\hline
                        {\bf base space:}     & $dP_6$-trans.~of $\IP^4[5]$ & new coords:  & $\tilde v = v^2$${}^\big.$ \\
                        involution:     & $v\mapsto -v$           &                   & $\tilde w= w^2$ \\
                        topology:       & $\chi(X) = -176$      & topology:         & $\chi(B)=-56$ \\
                        O7-planes:      & $D_v$ ($\chi=55$)       &                   & $D_{\tilde v}$ ($\chi=55$) \\
                                        & $D_w$ ($\chi=9$, $dP_6$)  &                   & $D_{\tilde w}$ ($\chi=9$, $dP_6$) \\
      prediction:     & $\chi^*(Y)=1728$        &                computation:  & $\chi^*(Y)=1728$${}_\big.$ \\ 
              \hline
                        involution:     & $x_2\mapsto -x_2$       & new coords: & $\tilde x_2=(x_2)^2$${}^\big.$ \\
                        topology:       & $\chi(X) = -176$           & topology:   & $\chi(B)=-65$ \\
                        O7-planes:      & $D_2$ ($\chi=46$)       &             & $D_{\tilde 2}$ ($\chi=46$) \\
      prediction:     & $\chi^*(Y)=612$         &          computation:   & $\chi^*(Y)=612$${}_\big.$ \\ \hline\hline
                \end{tabular}
        \caption{Two different orientifold involutions for the single del Pezzo transition of the quintic $\IP^4[5]$, which contains a $dP_6$ divisor. The first involution $v\mapsto-v$ is covered in detail in the main text.}
        \label{tab:upliftsdP6}
\end{table}

\begin{table}[ht]
        \centering
                \begin{tabular}{ll|ll}
                  \multicolumn{2}{c|}{orientifold data (3-fold)} & \multicolumn{2}{c}{uplift data (4-fold)${}_\big.$} \\ \hline\hline
                                        & & new coords:  & $\tilde v = v_1 v_2$${}^\big.$ \\
                  {\bf base space:} & $(dP_7)^2$-trans.~of $\IP^4[5]$ &                       & $\tilde w = w_1 w_2$ \\
                                        &                                  &                       & $\tilde h = v_1 w_1 + v_2 w_2$ \\
                        involution:     & $v_1\leftrightarrow v_2$         & topology:             & $\chi(B) = -48$ \\
                                        & $w_1\leftrightarrow w_2$         &                       & $D_{\tilde v}$ ($\chi=46$) \\
      topology:       & $\chi(X) = -152$                    &                       & $D_{\tilde h}$ ($\chi=24$) \\ 
      O7-planes:      & $D_{v_1}+D_{w_1}$ ($\chi=56$)    &                       & $D_{\tilde w}$ ($\chi=10$, $dP_7$) \\
      prediction:     & $\chi^*(Y)=1008$                 &               computation:        & $\chi^*(Y)=1008$${}_\big.$ \\ 
              \hline
                        involution:     & $u_2\mapsto -u_2$       & new coords: & $\tilde u_2=(u_2)^2$${}^\big.$ \\
                        topology:       & $\chi(X) = -152$           & topology:   & $\chi(B) = -58$ \\
                        O7-planes:      & $D_{u_2}$ ($\chi=36$)   &             & $D_{\tilde u_2}$ ($\chi=36$) \\
      prediction:     & $\chi^*(Y)=216$         &          computation:   & $\chi^*(Y)=216$${}_\big.$ \\ \hline\hline
                \end{tabular}
        \caption{Two different orientifold involutions for the double del Pezzo transition of the quintic $\IP^4[5]$, which contains two $dP_7$ divisors. The first involution (exchange of coordinates $v_1\leftrightarrow v_2$, $w_1\leftrightarrow w_2$) is covered in detail in the main text.}
        \label{tab:upliftsdP7sq}
\end{table}

\begin{table}[ht]
        \centering
                \begin{tabular}{ll|ll}
                  \multicolumn{2}{c|}{orientifold data (3-fold)} & \multicolumn{2}{c}{uplift data (4-fold)${}_\big.$} \\ \hline\hline
                        {\bf base space:} & $M_1^{(dP_8)}$ &              & ${}^\big.$ \\
                        involution:     & $x_3\mapsto -x_3$  & new coords:  & $\tilde x_3 = (x_3)^2$ \\
                        topology:       & $\chi(X) = -480$      & topology:    & $\chi(B) = -222$ \\
                        O7-planes:      & $D_3$ ($\chi=36$)  &              & $D_{\tilde 3}$ ($\chi=36$) \\
      prediction:     & $\chi^*(Y)=216$    &            computation:  & $\chi^*(Y)=216$${}_\big.$ \\ 
            \hline\hline
                        {\bf base space:} & $M_1^{(dP_9)}$ &              & ${}^\big.$ \\
                        involution:     & $x_3\mapsto -x_3$  & new coords:  & $\tilde x_3 = (x_3)^2$ \\
                        topology:       & $\chi(X) = -480$      & topology:    & $\chi(B) = -222$ \\
                        O7-planes:      & $D_3$ ($\chi=36$)  &              & $D_{\tilde 3}$ ($\chi=36$) \\
      prediction:     & $\chi^*(Y)=216$    &            computation:  & $\chi^*(Y)=216$${}_\big.$ \\ 
            \hline\hline
                \end{tabular}
        \caption{There are two different Calabi-Yau phases for the $dP_1$ surface, i.e.~the blowup of $\IP^2$ at a single point. Accordingly, one finds two phases for an elliptically-fibered threefold over $dP_1$ base and the respective fourfold uplift. The involution $x_3\mapsto - x_3$ corresponds to an inversion of a $dP_1$ base coordinate, see \cite{Blumenhagen:2008zz}.}
        \label{tab:upliftsM1}
\end{table}

\begin{table}[ht]
        \centering
                \begin{tabular}{ll|ll}
                  \multicolumn{2}{c|}{orientifold data (3-fold)} & \multicolumn{2}{c}{uplift data (4-fold)${}_\big.$} \\ \hline\hline
                  {\bf base space:} & $M_2^{(dP_8)^2}$        & new coords:  & $\tilde x_3=(x_3)^2$${}^\big.$ \\
                        involution:     & $x_3\mapsto -x_3$         &                   & $\tilde x_7=(x_7)^2$ \\
                        topology:       & $\chi(X) = -420$             & topology:  & $\chi(B) = -192$ \\
                        O7-planes:      & $D_3$ ($\chi=25, \Kb$)    &                   & $D_{\tilde 3}$ ($\chi=25$, $\Kb$) \\
                                        & $D_7$ ($\chi=11, dP_8$)   &                   & $D_{\tilde 7}$ ($\chi=11$, $dP_8$) \\
      prediction:     & $\chi^*(Y)=216$           &             computation:      & $\chi^*(Y)=216$${}_\big.$ \\ 
              \hline\hline        
                        {\bf base space:} & $M_2^{(dP_8,dP_9)}$        & new coords:  & $\tilde x_3=(x_3)^2$${}^\big.$ \\
                        involution:     & $x_3\mapsto -x_3$         &                   & $\tilde x_7=(x_7)^2$ \\
                        topology:       & $\chi(X) = -420$             & topology:  & $\chi(B) = -\frac{385}{2}=-192-\frac{1}{2}$ \\
                        O7-planes:      & $D_3$ ($\chi=24, \KK$)    &                   & $D_{\tilde 3}$ ($\chi=24$, $\KK$) \\
                                        & $D_7$ ($\chi=11, dP_8$)   &                   & $D_{\tilde 7}$ ($\chi=11$, $dP_8$) \\
      prediction:     & $\chi^*(Y)=378$           &             computation:      & $\chi^*(Y)=378$${}_\big.$ \\ 
              \hline\hline        
                        {\bf base space:} & $M_2^{(dP_9,dP_8)}$        & new coords:  & $\tilde x_3=(x_3)^2$${}^\big.$ \\
                        involution:     & $x_3\mapsto -x_3$         &                   & $\tilde x_7=(x_7)^2$ \\
                        topology:       & $\chi(X) = -420$             & topology:  & $\chi(B) = -\frac{383}{2}=-192+\frac{1}{2}$ \\
                        O7-planes:      & $D_3$ ($\chi=25, \Kb$)    &                   & $D_{\tilde 3}$ ($\chi=25$, $\Kb$) \\
                                        & $D_7$ ($\chi=12, dP_9$)   &                   & $D_{\tilde 7}$ ($\chi=12$, $dP_9$) \\
      prediction:     & $\chi^*(Y)=54$            &             computation:      & $\chi^*(Y)=54$${}_\big.$ \\ 
              \hline\hline        
                        {\bf base space:} & $M_2^{(dP_9)^2}$        & new coords:  & $\tilde x_3=(x_3)^2$${}^\big.$ \\
                        involution:     & $x_3\mapsto -x_3$         &                   & $\tilde x_7=(x_7)^2$ \\
                        topology:       & $\chi(X) = -420$             & topology:  & $\chi(B) = -192$ \\
                        O7-planes:      & $D_3$ ($\chi=24, \KK$)    &                   & $D_{\tilde 3}$ ($\chi=24$, $\KK$) \\
                                        & $D_7$ ($\chi=12, dP_9$)   &                   & $D_{\tilde 7}$ ($\chi=12$, $dP_9$) \\
      prediction:     & $\chi^*(Y)=216$           &             computation:      & $\chi^*(Y)=216$${}_\big.$ \\ 
              \hline\hline        
                        {\bf base space:} & $M_2^{(dP_{10})^2}$        & new coords:  & $\tilde x_3=(x_3)^2$${}^\big.$ \\
                        involution:     & $x_3\mapsto -x_3$         &                   & $\tilde x_7=(x_7)^2$ \\
                        topology:       & $\chi(X) = -420$             & topology:  & $\chi(B) = -\frac{383}{2}=-192+\frac{1}{2}$ \\
                        O7-planes:      & $D_3$ ($\chi=24, \KK$)    &                   & $D_{\tilde 3}$ ($\chi=24$, $\KK$) \\
                                        & $D_7$ ($\chi=13, dP_{10}$)&                   & $D_{\tilde 7}$ ($\chi=13$, $dP_{10}$) \\
      prediction:     & $\chi^*(Y)=54$            &             computation:      & $\chi^*(Y)=54$${}_\big.$ \\ 
              \hline\hline                      \end{tabular}
        \caption{There are five different Calabi-Yau phases for the $dP_2$ surface. Accordingly, one finds five phases for an elliptically-fibered threefold over $dP_2$ base and the respective fourfold uplift. The involution $x_3\mapsto - x_3$ corresponds to an inversion of a $dP_2$ base coordinate, see \cite{Blumenhagen:2008zz}. The half-integer Euler characteristics are related to the presence of an odd number of O3-planes in the respective cases.}
        \label{tab:upliftsM2}
\end{table}


\clearpage

\section{Tate algorithm}
\label{appb}
\begin{table}[h!]
        \centering
                \begin{tabular}{c|c|cc|cc@{${}\,\,\,\quad$}ccc}
                        sing.                        & discr.         & \multicolumn{2}{|c}{gauge enhancement}        & \multicolumn{5}{|c}{coefficient vanishing degrees}  \\
                        type                         & $\deg(\Delta)$ & type & group  & ${}\,\,\,\,\,\,a_1\quad{}$ & $a_2$ & $a_3$ & $a_4$ & $a_6{}_\big.$ \\ \hline\hline
                        $\mathrm{I}_0{}^\big.$               & 0      &            & ---          & 0 & 0 & 0     & 0     & 0 \\
                        $\mathrm{I}_1$                       & 1      &            & ---          & 0 & 0 & 1     & 1     & 1 \\
                        $\mathrm{I}_2$                       & 2      & $A_1$      & $SU(2)$      & 0 & 0 & 1     & 1     & 2 \\
                        $\mathrm{I}_3^{\,\mathrm{ns}}$       & 3      &            & [unconv.]    & 0 & 0 & 2     & 2     & 3 \\
                        $\mathrm{I}_3^{\,\mathrm{s}}$        & 3      &            & [unconv.]    & 0 & 1 & 1     & 2     & 3${}_\big.$ \\ \hline
                        $\mathrm{I}_{2k}^{\,\mathrm{ns}}$    & $2k$   & $C_{2k}$   & $SP(2k)$     & 0 & 0 & $k$   & $k$   & $2k$${}^\big.$ \\
                        $\mathrm{I}_{2k}^{\,\mathrm{s}}$     & $2k$   & $A_{2k-1}$ & $SU(2k)$     & 0 & 1 & $k$   & $k$   & $2k$ \\
                        $\mathrm{I}_{2k+1}^{\,\mathrm{ns}}$  & $2k+1$ &            & [unconv.]    & 0 & 0 & $k+1$ & $k+1$ & $2k+1$ \\
                        $\mathrm{I}_{2k+1}^{\,\mathrm{s}}$   & $2k+1$ & $A_{2k}$   & $SU(2k+1)$   & 0 & 1 & $k$   & $k+1$ & $2k+1$${}_\big.$ \\ \hline
                        $\mathrm{II}$                        & 2      &            & ---          & 1 & 1 & 1     & 1     & 1${}^\big.$ \\
                        $\mathrm{III}$                       & 3      & $A_1$      & $SU(2)$      & 1 & 1 & 1     & 1     & 2 \\
                        $\mathrm{IV}^{\,\mathrm{ns}}$        & 4      &            & [unconv.]    & 1 & 1 & 1     & 2     & 2 \\
                        $\mathrm{IV}^{\,\mathrm{s}}$         & 4      & $A_2$      & $SU(3)$      & 1 & 1 & 1     & 2     & 3 \\
                        $\mathrm{I}_0^{*\,\mathrm{ns}}$      & 6      & $G_2$      & $G_2$        & 1 & 1 & 2     & 2     & 3 \\
                        $\mathrm{I}_0^{*\,\mathrm{ss}}$      & 6      & $B_3$      & $SO(7)$      & 1 & 1 & 2     & 2     & 4 \\
                        $\mathrm{I}_0^{*\,\mathrm{s}}$       & 6      & $D_4$      & $SO(8)$      & 1 & 1 & 2     & 2     & 4 \\
                        $\mathrm{I}_1^{*\,\mathrm{ns}}$      & 7      & $B_4$      & $SO(9)$      & 1 & 1 & 2     & 3     & 4 \\
                        $\mathrm{I}_1^{*\,\mathrm{s}}$       & 7      & $D_5$      & $SO(10)$     & 1 & 1 & 2     & 3     & 5 \\
                        $\mathrm{I}_2^{*\,\mathrm{ns}}$      & 8      & $B_5$      & $SO(11)$     & 1 & 1 & 3     & 3     & 5 \\
                        $\mathrm{I}_2^{*\,\mathrm{s}}$       & 8      & $D_6$      & $SO(12)$     & 1 & 1 & 3     & 3     & 5${}_\big.$ \\ \hline
                        $\mathrm{I}_{2k-3}^{*\,\mathrm{ns}}$ & $2k+3$ & $B_{2k}$   & $SO(4k+1)$   & 1 & 1 & $k$   & $k+1$ & $2k$${}^\big.$ \\
                        $\mathrm{I}_{2k-3}^{*\,\mathrm{s}}$  & $2k+3$ & $D_{2k+1}$ & $SO(4k+2)$   & 1 & 1 & $k$   & $k+1$ & $2k+1$ \\
                        $\mathrm{I}_{2k-2}^{*\,\mathrm{ns}}$ & $2k+4$ & $B_{2k+1}$ & $SO(4k+3)$   & 1 & 1 & $k+1$ & $k+1$ & $2k+1$ \\
                        $\mathrm{I}_{2k-2}^{*\,\mathrm{s}}$  & $2k+4$ & $D_{2k+2}$ & $SO(4k+4)$   & 1 & 1 & $k+1$ & $k+1$ & $2k+1$${}_\big.$ \\ \hline
                        $\mathrm{IV}^{*\,\mathrm{ns}}$       & 8      & $F_4$      & $F_4$        & 1 & 2 & 2     & 3     & 4${}^\big.$ \\
                        $\mathrm{IV}^{*\,\mathrm{s}}$        & 8      & $E_6$      & $E_6$        & 1 & 2 & 2     & 3     & 5 \\
                        $\mathrm{III}^*$                     & 9      & $E_7$      & $E_7$        & 1 & 2 & 3     & 3     & 5 \\
                        $\mathrm{II}^*$                      & 10     & $E_8$      & $E_8$        & 1 & 2 & 3     & 4     & 5 \\
                        non-min                              & 12     &            & ---          & 1 & 2 & 3     & 4     & 6
                \end{tabular}
        \caption{Refined Kodaira classification resulting from Tate's algorithm. In order to distinguish the ``semi-split'' case $\mathrm{I}_{2k}^{*\,\mathrm{ss}}$ from the ``split'' case $\mathrm{I}_{2k}^{*\,\mathrm{s}}$ one has to work out a further factorization condition which is part of the aforementioned algorithm, see \S 3.1 of \cite{Bershadsky:1996nh}.} 
                \label{tab:TateTable}
\end{table}


\clearpage
\bibliography{rev28}
\bibliographystyle{utphys}


\end{document}